\documentclass{sig-alternate-05-2015}
\usepackage[utf8]{inputenc}
\usepackage{soul}
\usepackage{booktabs}
\usepackage{hhline}
\usepackage{color}
\usepackage{multicol}
\usepackage{graphicx}

\begin{document}
\makeatletter
\def\@copyrightspace{\relax}
\makeatother

\title{Analysis System for Theatrical Movie Releases Based on Movie Trailer Deep Video Representation} 

\author{
  Campo, Miguel\thanks{Corresponding author: miguelangel.campo-rembado@fox.com.  All authors are with 20th Century Fox.  Second author is also with Miso Technologies}\\
  \and
  Hsieh, Cheng-Kang\\
  \and
  Nickens, Matt\\
  \and
  Espinoza, JJ\\
  \and
  Taliyan, Abhinav
  \and
  Rieger, Julie\\
  \and
  Ho, Jean\\
  \and
  Sherick, Bettina\\
}


\maketitle   
\begin{abstract}

Audience discovery is an important activity at major movie studios. Deep models that use convolutional networks to extract frame-by-frame features of a movie trailer and represent it in a form that is suitable for prediction are now possible thanks to the availability of pre-built feature extractors trained on large image datasets. Using these pre-built feature extractors, we are able to process hundreds of publicly available movie trailers, extract frame-by-frame low level features (e.g., a face, an object, etc) and create video-level representations. We use the video-level representations to train a hybrid Collaborative Filtering model that combines video features with historical movie attendance records. The trained model not only makes accurate attendance and audience prediction for existing movies, but also successfully profiles new movies six to eight months prior to their release. 

\end{abstract}

\section{Introduction}
Video trailers are the single most critical element of the marketing campaigns for new films. They increase awareness among the general moviegoer population, communicate the plot of the movie, present the main characters, and reveal important hints about the story, the tone and the cinematographic choices.  They represent an opportunity for the filmmakers and the studio to learn customers preferences and to understand what aspects they liked or did not like. These insights typically determine the strategy for the rest of the marketing campaign. 

\begin{figure*}[h]
\centering
  \includegraphics[width=0.8\linewidth]{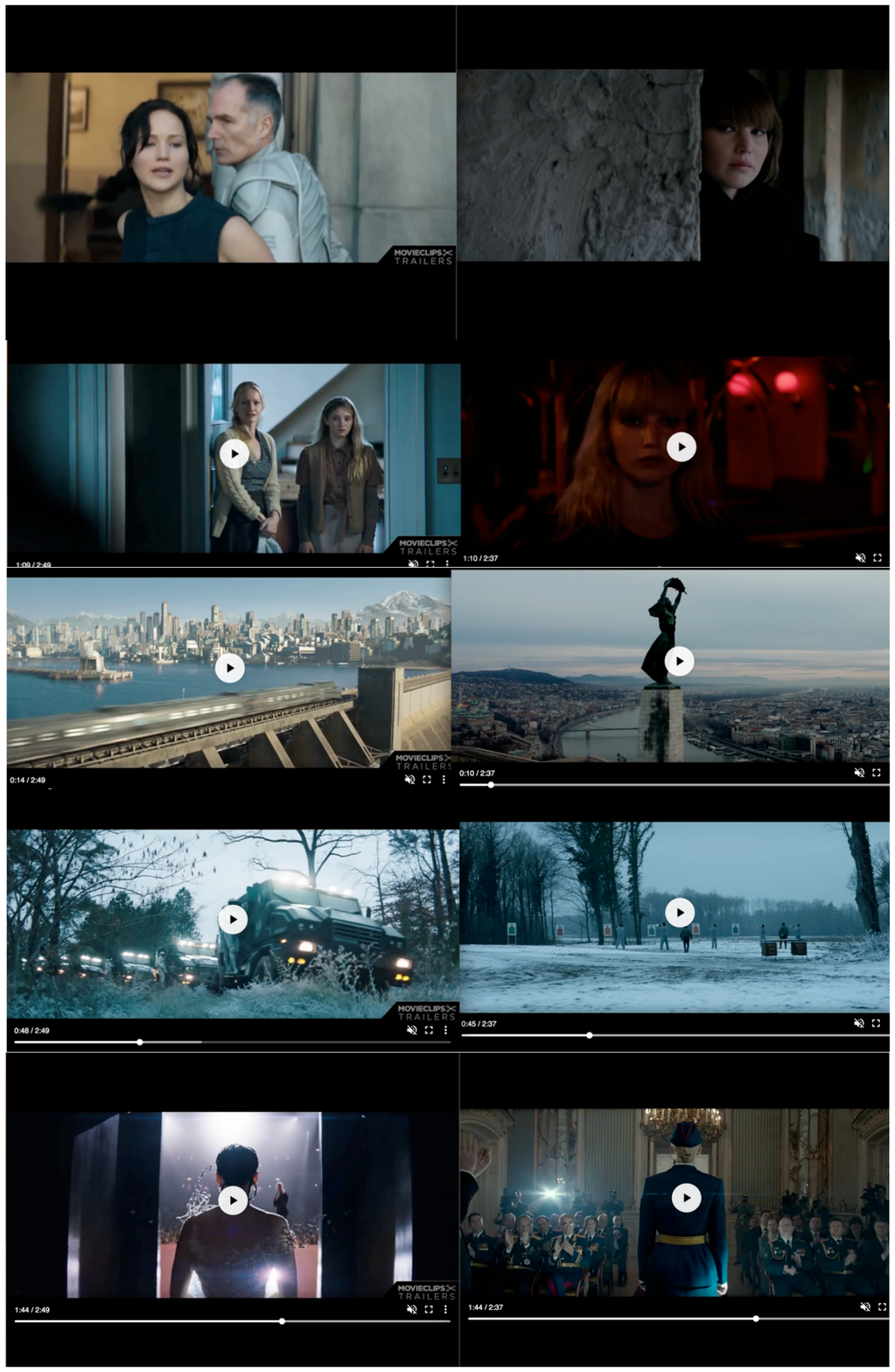}
  \caption{Left: The Hunger Games (Lionsgate); Right: Red Sparrow (20th Century Fox). }
  \label{fig:trailerexample}\vspace{-0.2cm}
\end{figure*}

In this paper we present a tool, which we call \textit{Merlin Video}, that uses computer vision to create dense representations of the movie trailer (see Figure \ref{fig:trailerexample}), and uses those representations to predict customer behavior. Movie studios have used computer vision before for video asset management and piracy detection. This is, to the extent of our knowledge, the first time that a studio uses low-level representation of movie trailers to extrapolate and predict customer interests.

The tool is based on a novel hybrid Collaborative Filtering (CF) model that captures the features of movie trailers, and combines them with attendance and demographic data to enable accurate in-matrix and cold-start recommendations. We evaluate the system using a large-scale dataset and observe up to 6.5\% improvement in AUC over various baseline algorithms that leverage text data (movie plot).  We show how a system that combines text and video inputs is used to assist real-world decision making at different stages of a marketing campaign.

The contribution of this paper is threefold: first, we present Merlin Video, the first of its kind recommendation system for movie theatrical releases designed specifically to tackle the challenges in cold-start and theatrical recommendations using movie trailer contents. Second, we carefully evaluate the performance of different variants of Merlin Video compared to baseline approaches, and show how Merlin Video can be used in a real-world decision making process. Finally, we discuss possible ways to combine text and video inputs to improve predictions in future research.

\section{Computer Vision applied to Audience Discovery}\label{sec:related}

Audience discovery is an important activity of major movie studios, especially for non-sequel films or for films that cross different genres. Audience discovery based on the analysis of the movie plot or script is useful because it helps surface and quantify strategic positioning options for the film that are deeply related to the development of the plot. Deep learning models that dissect a movie plot or movie script into smaller parts have been shown to successfully identify language repetitions that begin to capture aspects of human storytelling. Deep models that combine natural language analysis of the movie script with customer level data can find non trivial associations between the language patterns used in the plot description and actual customer behavior. One such model, Merlin Text, was described in \cite{merlintext} and illustrates the application of deep models and natural language processing to the audience prediction and discovery problem.

Once the movie is in production and video content becomes available (movie clips, teasers and trailer), most studios rely on customer research that is based on reactions to the video content. After a movie trailer is released, studios analyze online metrics to confirm pre-existing hypotheses about the nature and size of the audience, or to generate new ones. In the present, this inductive process is based on sophisticated analysis of a reduced number of data points that lends itself easily to human interpretation and linear storytelling. 

Analysis of movie trailers helps studios manage key messaging and strategic aspects of the marketing campaign. However, complex competitive dynamics cannot be captured or described by low dimensional analysis methods. Models of customer dynamics that incorporate contemporaneous competing options available in theaters or on streaming platforms are essential for effective understanding of audience choices. 

Deep learning models that analyze video content, like movie trailers, have recently become available thanks to creation of large video datasets and open source libraries. Movie recommendation systems that are based low-level video data have been described in \cite{videorecsys2015,videorecsys2016} and have proved to be effective at recommending movies with similar low level video features, like color or illumination. Deep models that use convolutional networks to extract low level video features have been shown to provide better representations of relevant video attributes and improve performance in the recommendation task \cite{computervision2015}. 

Moreover, deep models that use convolutional networks to extract frame-by-frame low level features (objects, landscapes, faces, etc) are now possible thanks to the use of pre-trained networks on vast image datasets \cite{youtube8m}. These models have been shown to generalize very well to other datasets thanks to the diversity and massive scale of the training data, which suggests the possibility of using these models in the context of movie trailers. Pre-trained models can be used to identify the features in the relevant frames of a video trailer. By finding a suitable representation of these features, and by feeding them to a model that has access to historical movie attendance records, it is possible to find non-trivial associations between the video trailer features, and future audiences choices after the movie releases in theaters or on streaming services.

\section{Movie Recommendation Systems}\label{sec:related_2}

Recommendation systems for movie theatrical releases are emerging machine learning tools used for greenlighting decisions, movie positioning studies, and marketing and distribution.  User level prediction, especially for mid-budget movies, is a key component of a successful multichannel distribution strategy (e.g. theater and streaming). Prediction is a challenging problem due to the inherent difficulty of modeling movies that have not been made, the zero-sum nature of competition, the distributed nature of the data ecosystem (which goes back to 1948's U.S. Supreme Court verdict forcing studios to divest from movie theater chains), and the difficulty of capturing cash transactions at the box office, esp. for younger audiences.  Prediction for frequent moviegoers, who bring a larger share of the theatrical revenues, is even harder as heavy moviegoers not only see more movies but also have a taste for different kind of movies.  

Predicting user behavior in \textit{pure cold-start} situations presents unique challenges, especially for items that are novel, such as non-sequels movies or movies that cross traditional genres. Moviegoers' attendance patterns are also markedly different from that of the online streaming users as discussed in Section \ref{sec:related}. Recently, Hybrid CF approaches, such as Collaborative Topic Regression and Factorization Machine, have been proposed to address the cold start problem in ratings predictions for scientific papers, online streaming, books, etc \cite{Rendle:2010:FM:1933307.1934620,wang2011collaborative}. The recent advance in deep neural network models also further extends the capability of these hybrid models and allows them to learn deeper insights from the content data. \cite{wang2015collaborative, volkovs2017dropoutnet, li2017collaborative, purushotham2012collaborative}.
In this paper we review Merlin Video, a recommendation system that uses a novel hybrid CF model to combine content information, in the form of a video trailer, with the historical movie attendance records. A previous version of Merlin that uses natural language plot descriptions is described in \cite{merlintext}.

The deep learning based hybrid CF model allows us to not only make accurate prediction for existing movies, but also profile new movies prior to their release or production. We evaluate Merlin Video's performance against state-of-the-art approaches and observe up to 6.5\% and 5.8\%  improvements in terms of AUC for in-matrix and cold-start movies respectively. We also present a real-world case study of how Merlin is used to assist the decision making process for movie positioning studies and audience analysis.

\section{Movie Attendance}\label{sec:related}
Movie recommendation for online streaming platforms has been well-studied in RecSys literature \cite{youtube, koren2009matrix, harper2016movielens}. However, little research has been done to study the recommendation and prediction problems for theatrical releases. Specifically, the study of the \textit{cold-start} prediction problem before and during movie production \cite{schein2002methods}.

The decision of going to a movie is different from that of consuming a streaming video. The former can be modeled with utility analysis in \textit{choices involving risk} \cite{friedman1948utility}, which captures well known differences between heavy and casual moviegoers.  Let $p$ denote the probability the user likes a movie, and assume that if she likes the movie, she experiences a utility increase of $U$, and if she does not, she experiences an utility decrease of $D$.  The expected change in utility, should she decide to go, is $pU+(1-p)D$, and the decision to go is optimal when $pU+(1-p)D)>0$ or $p>|D|/(U+|D|)$. For heavy users who go to movies regularly, the downside $|D|$ is low, and the chance that $p>|D|/(U+|D|))$ is high. As a result, they go to see many more movies.  For casual users, the downside can be rather large (especially when it is an expensive group activity). As a result, they will only go to those movies for which $p \approx 1$. This phenomena has profound impact on movie revenue and what movies get made.  

Movie-streaming differs from moviegoing. It is cheaper and reversible (one can stop the movie), and there is less risk involved. Casual streamers are not penalized for making mistakes when drawing from a broader catalog as casual moviegoers are if they do not like the movie.  

Collaborative filtering uses historical movie attendance records to infer movies and users' representations in a common space in which distance is a proxy for $p$, the probability the user likes a movie. The model proposed here identifies commonalities in the trailers to help determine that probability for each user.

\section{Methodology}

\begin{figure*}[h]
\centering
  \includegraphics[width=0.8\linewidth]{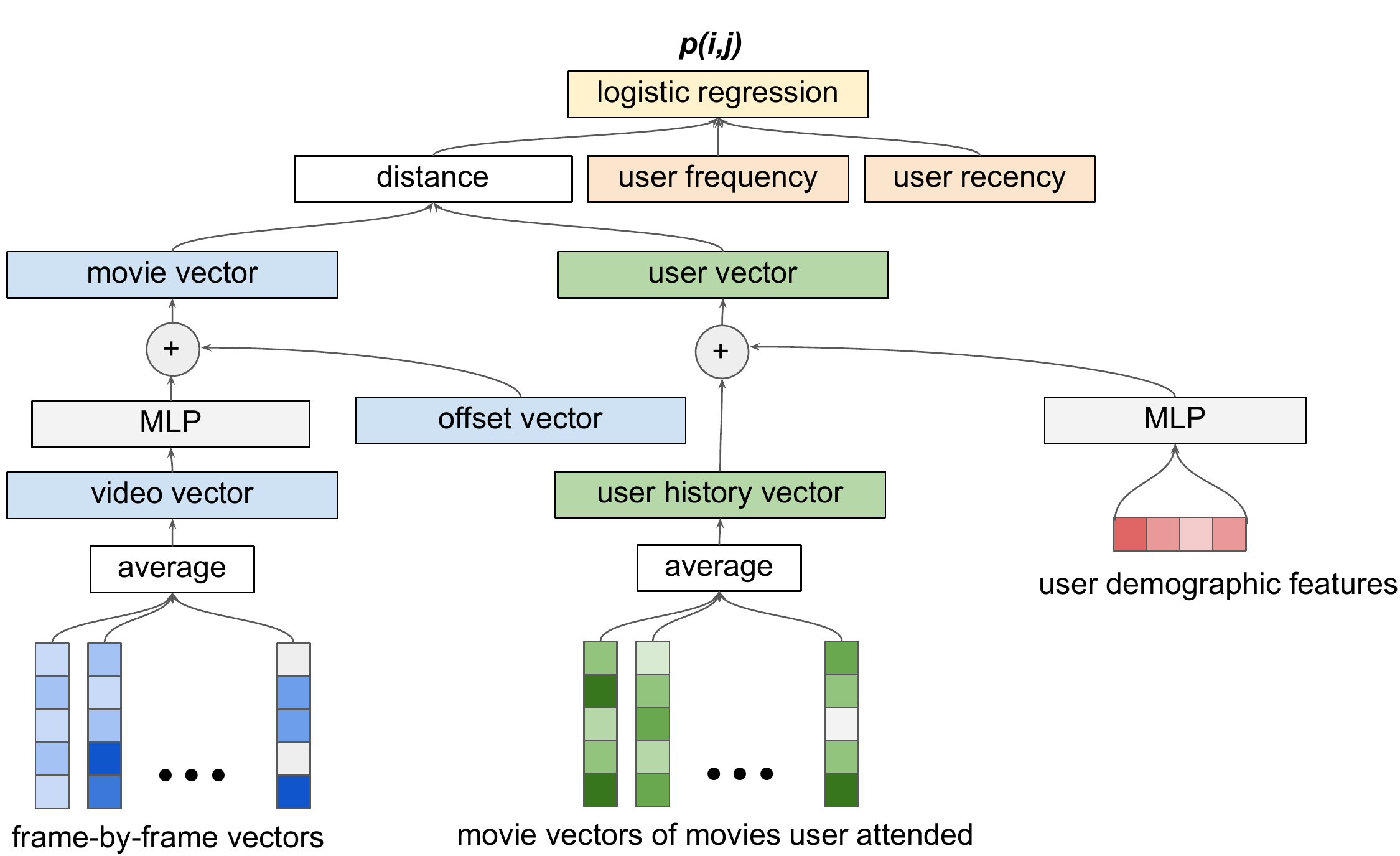}
  \caption{Overview of Merlin Video's hybrid recommendation model. A logistic regression layer combines a distance-based CF model with user's frequency and recency to produce the movie attendance probability. The model is trained end-to-end, and the loss of the logistic regression is backpropogated to all the trainable components. }
  \label{fig:overview}\vspace{-0.2cm}
\end{figure*}

In this section, we formally formulate the theatrical movie recommendation problem and present the design of Merlin Video. Given a set of movie $\mathcal{J}$ and a set of anonymous users $\mathcal{I}$ along with their movie attendance records, the goal of the system is to estimate the probability of a user $i$ to attend a movie $j$. Besides the attendance records, for each movie $j$, we have its video trailer denoted as $\mathbf{C}_j$; for each user $i$, we have his or her basic demographics information denoted as $\mathbf{D}_i$. These external information is used to not only improve the prediction accuracy for existing movies but also make possible the prediction for cold-start movies.

\subsection{Model Overview}
An overview of Merlin Video is illustrated in Figure \ref{fig:overview}. For each movie, we create a video vector from its video trailer (Section \ref{sec:content_vector}). Then, a multi-layer perceptron (MLP) \cite{ruck1990multilayer} is trained to project the video vector into a new embedding space. This space is shared by both movies and users and is where hybrid CF takes place. Each user in this space is represented by a user vector created from a fusion of the movie she attended and basic demographics information (Section \ref{sec:user_vector}). A distance-based hybrid CF framework is adopted to estimate the propensity between users and movies and encode that in their distance to each other. Finally, a logistic regression layer is trained to combine the movie-user distance with the user's movie attendance frequency and recency to produce the final user attendance probability to the movie in question (Section \ref{sec:freq}). To allow efficient information flow, the whole model is trained end-to-end: i.e. the loss from the logistic regression is backpropagated to each trainable component in the model. In the following, we describe main components in Merlin Video and their design rationale.

\subsection{Video Vector}\label{sec:content_vector}
For each movie $j$, we compute the video embedding of the movie denoted by $\mathbf{x}_j$. 

We compute the embedding using pre-trained models (we use YouTube-8M), a large-scale labeled video dataset. Specifically, we use the publicly available YouTube8M Feature Extractor to extract features at the frame-level at a 1-second resolution. 

The frame-level features are extracted using Inception-V3 image annotation model, pre-trained on ImageNet. Each frame generates a 1024-long dense vector.  We retain the first 100 frames and create a vector of video features by averaging the frame vectors. The video-level vector is generated for hundreds of trailers publicly available in YouTube. 

For this process, we use a cloud hosted virtual machine with 16 CPUs and 60 GB of memory to support and processing hundreds of videos (638,000+ video frames in total) using the pre-trained model weights (100 Megabytes) from the Inception Model.

\subsection{Movie Vector}\label{sec:movie_vector}
The video vector $\mathbf{x}_j$ captures the low-level features of the video trailer. We project $\mathbf{x}_j$ into to a movie embedding space using a multi-layer perceptron (MLP) $f$ (i.e. $\mathbf{v}_j = f(\mathbf{x}_j)$). After training, the MLP will learn to transform the unsupervisedly-learned video vector into a representation that is more aligned with the actual movie attendance records. For example, the projection may amplify the features suggesting a action scene and suppress other less relevant features in the video vectors. For the same reason, an offset vector $\mathbf{\theta}_j$ is learned for each movie $j$ to capture additional information in attendance records. The final movie vector $\hat{\mathbf{v}}_j$ is set to be the sum of the initial projection and the offset vector, i.e. $\hat{\mathbf{v}}_j = \mathbf{v}_j + \mathbf{\theta}_j$. 

From a Bayesian perspective, the initial projection $\mathbf{v}_j = f(\mathbf{x}_j)$ can be seen as a \textit{prior} of a movie given its content, and the offseted version of it $\hat{\mathbf{v}}_j$ is the \textit{posterior}. The offset vector $\mathbf{\theta}_v$ captures the information that is not available in the low level video representation but present in the actual movie attendance records after the movie release. 

For a movie in production, where no attendance information is available, the movie vector is set to be its initial projection $\mathbf{v}_j$. This, however, does not mean the offset is not useful to the cold-start prediction. On the contrary, we observe that the offset vector allows more flexibility in modeling content information, and helps prevent outliers, e.g. a blockbuster movie whose performance may have little to do with its video trailer, from negatively affecting the content modeling of other non-blockbuster movies (as its effect is absorbed by the offset). Similar results have been discussed in the previous regression based models for scientific papers and other recommendation tasks \cite{wang2015collaborative, wang2011collaborative, hsieh2017collaborative}.

\subsection{User Vector}\label{sec:user_vector}
On the user side, for each user $i$, we create a user vector from her movie attendance record and demographics data. Specifically, for a user $i$, we first take the average of the movie vectors of the movies she attended to create a user history vector, i.e. $\mathbf{h}_i = \frac{1}{|\mathcal{S}_i|}\sum_{j \in \mathcal{S}_i}{\hat{}{\mathbf{v}}_j}$, where $\mathcal{S}_i$ is the set of movies user $i$ attends. During the training, the loss will be backpropagated to all the movie vectors in the set $\mathcal{S}_i$. This setup is similar to the one proposed in \cite{youtube}. The rationale behind this more restrictive representation is that there are significantly more moviegoers than movies. Learning a unique user vector for each user will 1) take significantly longer time to train and 2) suffer from over-fitting due to relatively small number of movie attendances per user. 

To incorporate the user demographic information, we define another MLP $g$ to project one-hot-encoded demographic information $\mathbf{D}_i$ into the same space as the user history vector. The final user vector is defined as $\hat{\mathbf{u}}_i = \mathbf{h}_i + g(\mathbf{D}_i)$. Note that, an alternative is to concatenate the demographic features $\mathbf{D}_i$ with the history vector $\mathbf{h}_i$ before applying MLP \cite{youtube}. Empirically, we found our approach produces better performance in our use case as it allows more direct information flow between the movie vectors.

\subsection{Collaborative Metric Learning}
With the movie vectors and user vectors defined, the next step is to define their \textit{interaction function} to estimate the user-item propensity \cite{yang2018openrec}. Collaborative metric learning (CML) \cite{hsieh2017collaborative} is adopted to estimate the propensity. CML encodes user-item propensity in their \textit{euclidean distance} such that a movie will be closer to the users who attend it and relatively further away from the users who do not. The use of euclidean distance makes CML more effective in capturing fine-grained characteristics of items and user and is particularly well-suited for our use case. 

\subsection{Preference, Frequency and Recency}\label{sec:freq}
So far, we have established a model that estimates the propensity between users and movies. However, as a well-known phenomena in movie industry and market research, consumers' consumption frequency and recency play a major role in their consumption patterns. To incorporate these factors, we use a logistic regression layer that combines 1) the user-movie distance defined earlier, 2) the attendance frequency, and 3) the attendance recency of user $i$. Our experiment shows frequency and recency are strong predictor to the attendance probability. 

However, when training the model end-to-end, one issue early on is that the model will choose to over-fit these factors and ignore the distance factor, whose values are mostly random at the beginning of the training. Therefore, a heavy dropout ($p=0.5$) is applied to the logistic layer to ensure the model to incorporate every factor. The final logistic regression loss is backprogated to all the trainable components, including offset vectors $\mathbf{\theta}_j$ and the parameters of MLP $f$ and $g$.

\section{Performance Evaluation}
We evaluate Merlin Video's performance using a double-blind anonymized, user privacy compliant, movie attendance dataset that combines data from different sources with hundreds of movies released over the last years, and millions of attendance records. We hold out the attendance records of the most recent 50 movies for cold-start evaluation. We hold out 10\% of the remaining records for validation and another 10\% for testing. All the models are trained until convergence on the validation set and evaluated on the testing set. 

The model is trained using stochastic gradient descent with mini-batches. Every batch contains an even mix of positive and negative user-item samples. During evaluation, for each positive user-item pair, we sample nine users that did not go to the movie to make positive-negative ratio aligned with the average movie attendance rate. 

\subsection{Evaluation Results}
We compared Merlin Video's performance with our previous plot-based model, Merlin Text \cite{merlintext}, as well as, the following baseline models: 1) Recency-Frequency (\textbf{RF}), a simple model that applies logistic regression on only user frequency and recency to determine the attendance probability. This is a popular prediction method in consumer and market research. 2) Probabilistic Matrix Factorization (\textbf{PMF}), which is a well-known (non-hybrid) latent vector based CF model. PMF cannot perform cold-start recommendations. 3) Collaborative Deep Learning (\textbf{CDL}), the state-of-the-art hybrid CF model that co-trains a neural network with PMF to integrate the content information. To have a fair comparison, we replace the autoencoder in CDL with the MLP we used in Merlin, which shows better performance in modeling text-based content (movie plots). 

We use Area-Under-Curve (AUC) \cite{hanley1982meaning} as our performance metric. Note that, since Merlin Video's goal is to tell what kind of moviegoers a movie will attract, and provide insights for movie production, the more commonly used user-level ranking metrics, such as Top-k recall, are less applicable here. 

Table \ref{table:perf} summarizes the evaluation results. As shown, our previous model Merlin Text has up to 12.2\% and 7.1\% improvements over the best baseline algorithm for in-matrix and cold-start scenarios respectively. While CDL performs the best among the text-based baseline algorithms, RF model, despite its simplicity, also shows competitive accuracy particularly in the cold-start scenario.  

Table \ref{table:perf} also shows the performance of Merlin Video using the same basic CF architecture, but substituting text data (movie plots) by video data (movie trailer). As shown, the use of offset vector is also beneficial to the in-matrix scenario, although to a lesser extent. The AUC performance of Merlin Video is better than the performance of text-based baseline models. Out-of-matrix AUC performance for Merlin Video is below the performance of Merlin Text. This is to be expected since the architecture of the hybrid CF model was fine tuned to the text case, not the video case.

\begin{table}[t]
\centering
\label{table:perf}
\begin{tabular}{|l|c|c|}
\cline{1-3}
\textbf{Model} & \textbf{In Matrix} & \textbf{Cold Start} \\ \hhline{===}
RF          &   0.717            &  0.676   \\
PMF          &    0.701               &       NA              \\
CDL (Text)            &      0.720         &    0.669        \\ \cline{1-3}
Merlin (Text)         &      \textbf{0.807}      &    \textbf{0.724}               \\ \cline{1-3}
Merlin Video         &      \textbf{0.747}      &    \textbf{0.708}     \\ \cline{1-3}
\end{tabular}
\caption{AUC of baselines and Merlin variants}\vspace{-0.2cm}
\end{table}


\begin{figure*}[t]
  \centering
\includegraphics[width=0.8\linewidth]{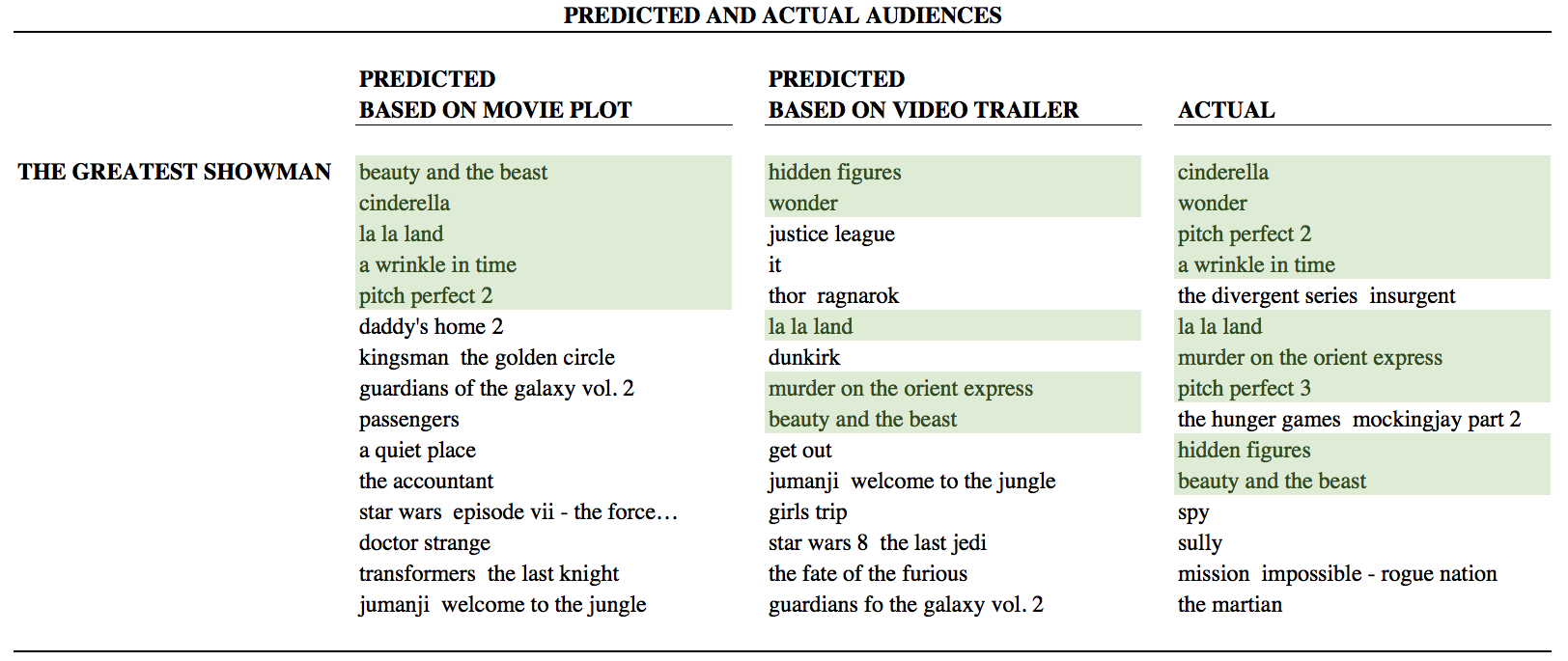}
  \caption{Movie Comp Table Generated By Merlin Text and Merlin Video}
  \label{fig:comp}\vspace{-0.3cm}
\end{figure*} 

\subsection{Use Case: Movie Comp Analysis}
Being able to predict accurate audience composition in terms of comparable movies is a valuable capability for movie studios to plan franchises, produce successful movies, optimize release windows, and execute on-target marketing campaigns. We use Merlin Video to identify the users that have the highest predicted probability to see the movie based on the trailer alone, and then identify other movies (i.e. comps) that were seen by those users. To evaluate the accuracy of these comps, once the movie is released in the theaters, we go back and compare the predicted list with the actual list. 

Figure \ref{fig:comp} shows the predicted (pre-release) and actual list of comparable movies for The Greatest Showman, a family/musical title. We have highlighted the movies (or sequels) that were accurately predicted based on top-N criteria. As seen from the table, predictions based on the video trailer (second column) are different from predictions based on the plot description (first column). For example, the audience of the movie Hidden Figures is very likely to attend the movie The Greatest Showman based on the video trailer alone, but not based on the plot description. There are however some similarities between the text-based and video-based predicted lists. For example, the audience of the movie Beauty and the Beast is very likely to attend based on movie plot analysis, and likely to attend based on video trailer analysis.


\section{Discussions and Future Work} \label{sec:graph}
Text-based and Video-based predictions use different information, as shown in the example above. The offset movie vectors learned by the collaborative filter depend on the type of content information (text or video) that we use to train the model.  Moreover, there are more movies highlighted in the third column of figure \ref{fig:comp} than in the first or second columns independently, which means that the two predictions complement each other and that we could benefit from building hybrid models that use text and video information.

Our results are based on mean-pooled video-level features that are fed directly to the collaborative filter network. Although this facilitates model training, mean-pooling frame topics results in severe loss of temporal information (e.g., does the car chase happen before or after the explosion?).  In addition, the fact that text-based and video-based models both generate high-quality but different comparison tables in Figure \ref{fig:comp} hints that they might reflect different aspects of a movie. We are currently exploring different models that can combine textual plot data with frame-by-frame features in order to create video vectors that begin to capture a suitable representation of the underlying video story.

\section{Conclusions}
We used mean-pooled video level features of movie trailers and movie attendance data to train a hybrid Collaborative Filtering model. The CF model was used to predict customer behavior six to eight months after the trailers became available. The out-of-matrix performance of the video model was comparable to the performance of the text-based model. Temporal modeling approaches for movie trailers that identify appropriate video representations and that take into account event sequence or higher order constructs like cinematography and visual storytelling are expected to achieve significantly better results than temporal pooling.


 \section{Acknowledgments}
The authors would like to thank Scott Bishoff, Kimberly Flaster, Ashley Cartwright and Jessica Gray from 20th Century Fox, and Vijay Reddy from Google, for their feedback.

\bibliographystyle{abbrv}
\bibliography{bib}  
%
%

\end{document}